\documentclass[twocolumn,aps,superscriptaddress,showpacs,nofootinbib,floatfix]{revtex4}

\usepackage{bm,feynmf}
\usepackage{amsmath,amssymb}
\usepackage{graphicx}
\usepackage[normalem]{ulem}  
\usepackage[dvips]{color} 

\renewcommand\sout{\bgroup \color{red} \ULdepth=-.5ex \ULset}

\begin{document}
\title{Jet Fragmentation via Recombination of Parton Showers}
\author{Kyong Chol Han}\email{khan@comp.tamu.edu}
\affiliation{Cyclotron Institute and Department of Physics and Astronomy, Texas A$\&$M University, College Station, TX 77843-3366, USA}
\author{Rainer J. Fries}\email{rjfries@comp.tamu.edu}
\affiliation{Cyclotron Institute and Department of Physics and Astronomy, Texas A$\&$M University, College Station, TX 77843-3366, USA}
\author{Che Ming Ko}\email{ko@comp.tamu.edu}
\affiliation{Cyclotron Institute and Department of Physics and Astronomy, Texas A$\&$M University, College Station, TX 77843-3366, USA}

\begin{abstract}
We propose to model hadronization of parton showers in QCD jets through a hybrid
approach involving quark recombination and string fragmentation. This is
achieved by allowing gluons at the end of the perturbative shower evolution to
undergo a non-perturbative splitting into quark and antiquark pairs, then
applying a Monte-Carlo version of instantaneous quark recombination, and
finally subjecting remnant quarks (those which have not found a recombination
partner) to Lund string fragmentation. When applied to parton showers from the
PYTHIA Monte Carlo event generator, the final hadron spectra from our calculation compare quite well to PYTHIA jets that have been hadronized with the
default Lund string fragmentation. Our new approach opens up the possibility
to generalize hadronization to jets embedded in a quark gluon plasma.
\end{abstract}

\maketitle
\section{Introduction}

Hadron production from jets in high-energy collisions of hadrons or nuclei is
often parameterized through fragmentation functions, using the universality of
the process as given by factorization theorems of quantum chromodynamics (QCD)~\cite{Collins:1981uw}. On a
microscopic level, hadron production in jets can be modeled very well through
a perturbative evolution of the parton shower inside the jet using DGLAP
splitting kernels to some low virtuality cutoff $Q_0$, followed by a non-perturbative hadronization model like the Lund string model or cluster hadronization applied to the parton shower. Event generators like PYTHIA~\cite{Sjostrand:2006za} and HERWIG~\cite{Corcella:2002jc} have successfully implemented such strategies to describe high momentum hadron production in $e^{+}+e^-$, $p+p$ and other processes.

In collisions of heavy nuclei at high energy, QCD factorization in jet 
hadronization is broken up to much higher hadron momentum, roughly 6-8 GeV/$c$ 
at typical collider energies, compared to the situation in elementary $e^{+}+e^-$ and $p+p$
collisions. This can be readily seen from the baryon enhancement measured in
nuclear collisions both in Au$+$Au collisions at the Relativistic Heavy Ion
Collider (RHIC)~\cite{Adler:2003qi} and in Pb$+$Pb collisions at the Large
Hadron Collider (LHC)~\cite{Abelev:2012wca}. It has been suggested that hadron
production at intermediate momenta, i.e. 2-8 GeV/$c$, can be described through
the process of quark recombination or
coalescence~\cite{Greco:2003xt,Fries:2003vb,Greco:2003mm,Fries:2003kq,Fries:2004ej,Fries:2008hs}.
It is an intriguing idea to combine the concepts of quark recombination and
parton showers since recombination can be easily generalized to the
hadronization of jets in dense environments as found in relativistic heavy ion
collisions.  In fact, quark recombination was applied to hadronization in jets
in the early days of QCD~\cite{Das:1977cp,Chang:1980kh,Migneron:1981cv}, and
more recently by Hwa and Yang~\cite{Hwa:2003ic}. However, parton showers in
those early works were not obtained from the sophisticated parton Monte Carlo
generators available today, but rather fitted to data or determined from specific models. In addition, earlier work also used event-averaged spectra, ignoring fluctuations coming from the small number of partons in each jet.

Here we show that essential aspects of hadron production in jet showers can be reproduced if we replace Lund string fragmentation in PYTHIA with an improved recombination model. We work with quarks and gluons at the end of their perturbative shower evolution, then let gluons decay into quark-antiquark pairs, evaluate quark recombination probabilities based on hadron Wigner functions by Monte Carlo sampling, and finally reapply Lund string fragmentation to those quarks which have not found a recombination partner. Finally, we compare our results to full PYTHIA results which simply hadronize entire showers by string fragmentation.

The paper is organized as follows. In the next section we describe how we
prepare perturbative parton showers and extract the constituent quark
distributions in phase space. In Sec.\ III, we describe the recombination
model used in the present study and our treatment of remnant partons. In
Sec.\ IV, we discuss our results and compare to full PYTHIA with string
fragmentation. We conclude in Sec. V.  Also included is an Appendix to derive
the recurrence relation for the overlap integral between Gaussian wave packets
and harmonic oscillator wave functions in the Wigner formalism used in
the recombination calculation. 
Although in this work we deal strictly with jets in the vacuum, our motivation 
derives from the desire to generalize our approach to jets in a QCD medium 
later on \cite{Han:new}.

\section{Parton Showers}

\begin{figure}[tb]
\centerline{\hspace{0.4cm}\includegraphics[width=\columnwidth]{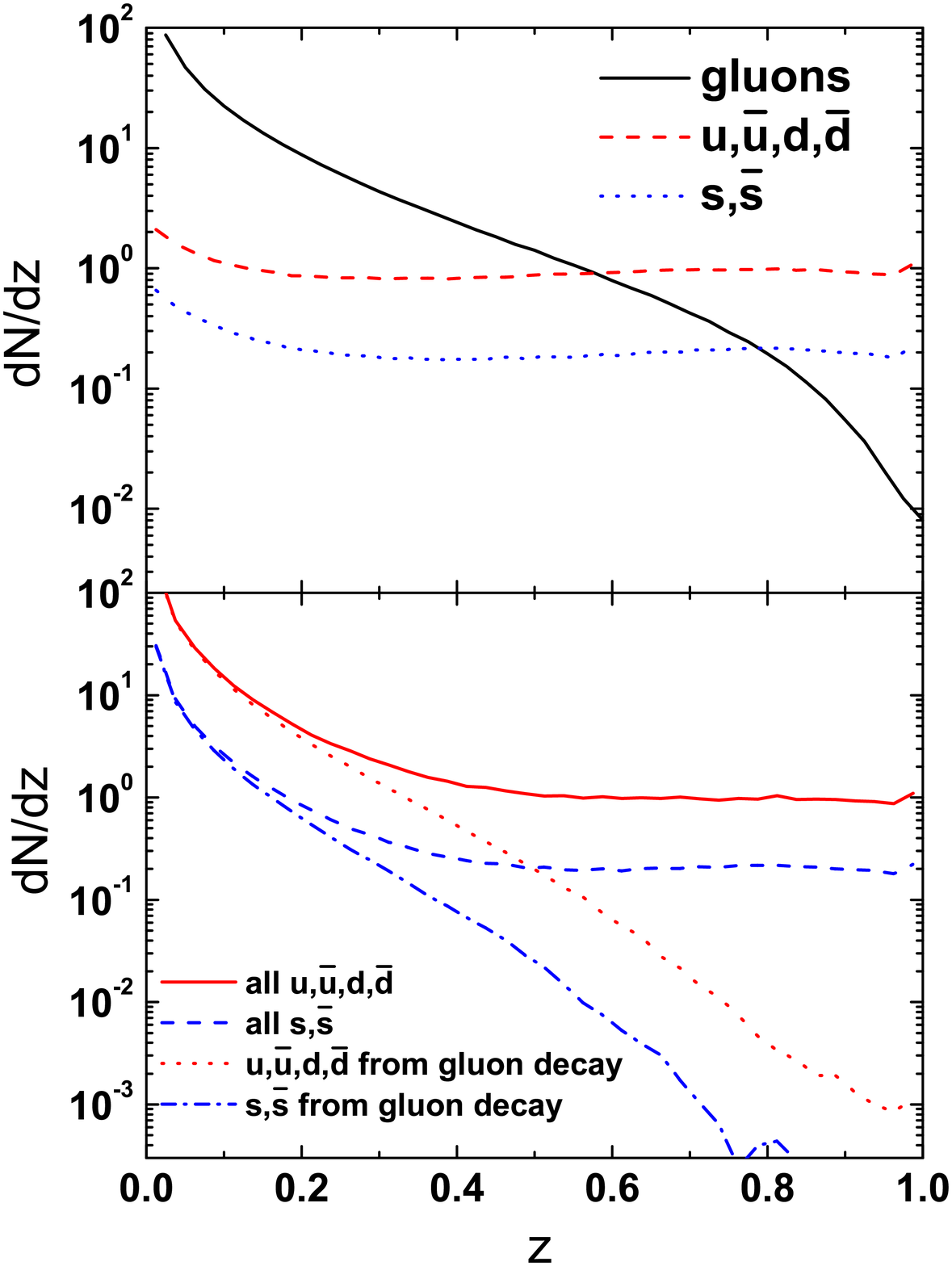}}
\caption{(Color online) Distribution $dN/dz$ of shower partons in terms of the momentum fraction $z$ of the initial jet momentum at the end of the perturbative shower evolution for a jet of 100 GeV before (upper panel) and after forcing gluon decays into quark-antiquark pairs (lower panel).}
\label{gluondecaygraph}
\end{figure}

\begin{figure}[tb]
\centerline{\hspace{0.4cm}\includegraphics[width=\columnwidth]{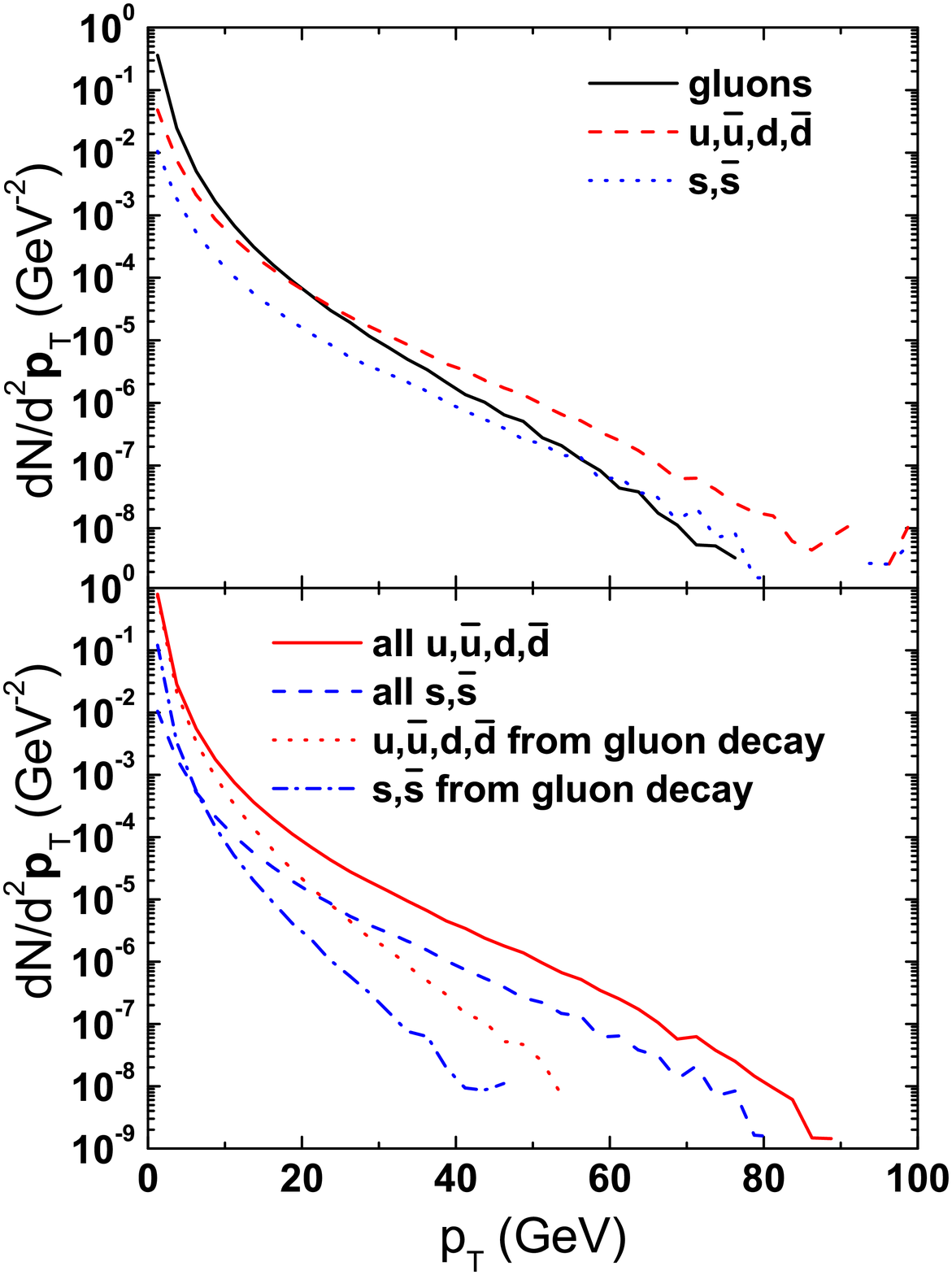}}
\caption{(Color online) Same as Fig.\ \ref{gluondecaygraph} for 
the shower parton transverse momentum distribution $dN/d^2 {\bf p}_T$.}
\label{gluondecaygraph2}
\end{figure}

We are not concerned here with the mechanisms involved in creating parton
showers. We use PYTHIA 6.3~\cite{Sjostrand:2006za} as a tool to create
perturbative parton showers as input to our hadronization procedure. 
PYTHIA 6.3 also serves as our benchmark for hadronization when we run pure
Lund string fragmentation on the same ensemble of parton showers.
Of course, another event generator that allows the extraction of shower partons
from a jet before hadronization would work as well.
Unless explicitly stated otherwise, the results presented here use
monoenergetic jets of energy 100 GeV which are extracted from $e^{+}+e^{-}$
collisions at a center-of-mass energy of $\sqrt{s} = 2 E_\mathrm{jet}=200$ GeV
in PYTHIA 6.3. By setting the cutoff for the perturbative evolution of  the
jet to $Q_0 = 1$ GeV, we extract the final parton configuration before string
breaking. The upper panels of Figs.~\ref{gluondecaygraph} and \ref{gluondecaygraph2} show
the resulting light quark ($u,d,\bar u, \bar d$), strange quark ($s,\bar s$) and gluon
($g$) spectra as functions of their longitudinal momentum fraction $z$ in the 
jet and as functions of their momentum $p_T$
transverse to the jet axis, respectively. More precisely we define
\begin{eqnarray}
z =\frac{\mathbf{p} \cdot \mathbf{P}_{\mathrm{jet}}}{|\mathbf{P}_{\mathrm{jet}}|^2},\quad
p_T  = \frac{\sqrt{|\mathbf{p}|^2 |\mathbf{P}_{\rm jet}|^2-({\bf p}\cdot{\bf P}_{\rm jet})^2}}{|\mathbf{P}_{\rm jet}|}.
\end{eqnarray}
where $\mathbf{p}$ is the 3-momentum of the considered parton and
$\mathbf{P}_{\mathrm{jet}}$ is the 3-momentum of the original parton creating
the jet. The spectra $dN/dz$ and $dN/d^2 {\bf p}_T$ are for one jet averaged over an ensemble of $10^6$ PYTHIA jets with $E_\mathrm{jet} = 100$ GeV.

Since recombination models are usually built on the premise of dominance of
the lowest Fock states in hadron wave functions, similar to hadronization in
exclusive processes~\cite{Lepage:1979zb,Lepage:1980fj}, only quarks and
antiquarks are considered (see Ref.~\cite{Muller:2005pv} for a study on higher
Fock states). Successful recombination models therefore postulate a
(non-perturbative) splitting of gluons into quark-antiquark pairs. Using
constituent quarks with masses $m_{u,d}=0.33~\rm{GeV}$ for light quarks and
$m_s=0.5~\rm{GeV}$ for strange quarks, consistent with PYTHIA, we thus let
gluons at the end of their perturbative evolution decay with their remaining
virtualities between $2m_{u,d}$ and $m_\mathrm{max}$, where $m_\mathrm{max}>
2m_s$ should be of the order of the scale $Q_0$. $m_\mathrm{max}$ is in
principle a parameter and its value will mostly influence the ratio of strange
to non-strange hadrons. We set this parameter to 1.25 GeV throughout this
work. We decay gluons isotropically in their rest frame into $q\bar q$ pairs. The decay chemistry gives equal weight to $u\bar u$ and $d\bar d$ pairs for gluon virtualities between $2 m_{u,d}$ and $2m_s$, while above the strangeness threshold the ratio of light to strange quarks is simply given by phase space and the vector nature of the decay as
\begin{equation}
  \frac{\Gamma(g^*\to u\bar u, d\bar d)}{\Gamma(g^* \to s\bar s)} = 2
  \frac{m^2 +2m^2 _{u,d}}{m^2 +2m^2 _s} \sqrt{\frac{m^2 -4m^2 _{u,d}}{m^2 -4m_s ^2}}.
\end{equation}
We do not consider heavy quarks in this study.

In PYTHIA, the final virtuality of shower gluons is forced to zero when the
value becomes smaller than $Q_0$. While one could in principle undo this step,
it turns out to be sufficient to reintroduce the non-perturbative gluon
virtuality manually without rebalancing momenta in the last splitting. We find
the typical error in total energy of the shower introduced this way is less
than 1\% for 100 GeV jets. The lower panels of Figs.~\ref{gluondecaygraph} and
\ref{gluondecaygraph2} show
the spectra of light and strange quarks from gluon decays for the same sample
of 100 GeV jets used previously, together with the total light and strange quark
spectra. The average number of quark and antiquarks in these 100 GeV jet
showers after decays is about 13. 

In principle, quark recombination could be formulated completely in momentum
space (see Ref.~\cite{Han:2012hp} for applications to jet showers). However,
for future applications in heavy ion collisions, where thermal partons will 
have nontrivial space-momentum correlations, we espouse a formulation of
quark recombination employing Wigner functions with both momentum and
space-time information. We are therefore led to introduce a space-time
structure of showers. We do this based on two simple premises: (i) Virtual
partons with virtuality $Q$ have an average life time $1/Q$ in their rest
frame before splitting. This time is then properly boosted into the lab
frame. (ii) The centers of wave-packets representing partons move on free
classical trajectories given by the velocity $\mathbf{p}/E$ of the parton in
the lab frame, where $E$ is the parton energy. 

\begin{figure}[tb]
\centering
\includegraphics[width=\columnwidth]{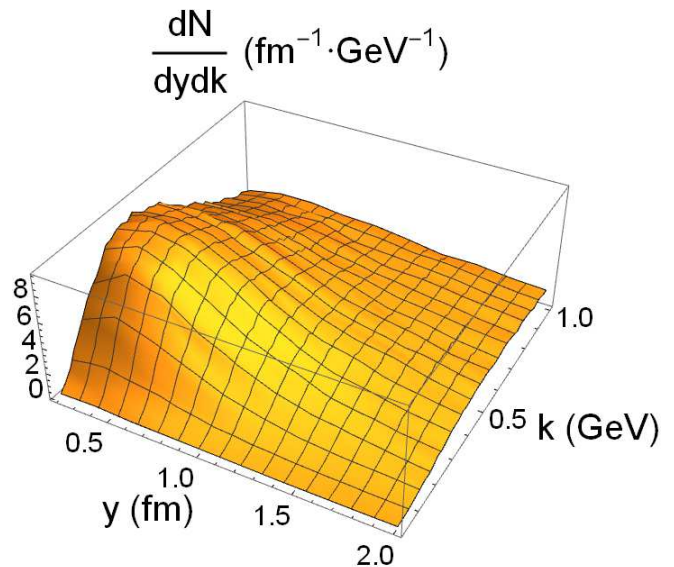}\\
\caption{(Color online) Statistical distribution of quark-antiquark pairs 
  in 100 GeV jet parton shower in terms of relative spatial and momentum coordinates
  $y$ and $k$ of the pair.  The coordinates are
  defined in the common rest frame of the pair at the time the latter parton 
  is created in the shower. }
\label{drdkMB}
\end{figure}

In the jet rest frame the spatial density of its shower partons depends on the 
time they are produced. However their density in momentum space is about 
0.025/GeV$^{3}$ and significantly smaller than the corresponding value of about 2.5/GeV$^{3}$ for partons in a quark-gluon plasma at its phase transition temperature. One can analyze this parton initial
state for hadronization more quantitatively.
As we will discuss in detail in the next section, the decisive physical
quantities for recombination between a particular quark and antiquark pair 
to occur are the relative distances $y$ and $k$ between 
the partons in space and momentum space measured at a common time in the rest
frame of the pair.  In Fig.~\ref{drdkMB} we show the statistical distribution of all 
quark-antiquark pairs we find in 100 GeV jet parton showers (normalized to 
one jet) as a function of their distances $y$ and $k$ in 
their common rest frame at the time when the latter parton of a pair is created.
We find that this distribution peaks at $y\sim 0.5$ fm and
$k\sim 0.3$ GeV, although large tails exist. This points to the 
existence of a ``bulk'' of partons in a jet shower which are quite close in
phase space and amenable to recombination, while another, non-negligible fraction
of partons will be far removed from other partons in phase space.

\section{Quark Recombination}

Instantaneous quark recombination is most conveniently expressed in terms of
an overlap of Wigner functions \cite{Fries:2008hs}. The momentum distributions of mesons and baryons formed from recombination of quarks are generally given by
\begin{eqnarray}
\label{Eq1}
\frac{dN_M}{d^3 {\mathbf P}_M} &=& g_M\int d^3{\bf x}_1 d^3{\bf p}_1 d^3{\bf x}_2
d^3{\bf p}_2 f_{q}({\bf x}_1, {\bf p}_1) f_{\bar{q}}({\bf x}_2, {\bf p}_2)\nonumber\\
&&\times W_{M}({\bf y}, {\bf k})\delta^{(3)}({\bf P}_{M}-{\bf p}_1 -{\bf p}_2) \, ,
\end{eqnarray}
and
\begin{eqnarray}
 \label{Eq2}
\frac{dN_B}{d^3 {\bf P}_B} &=& g_B\int d^3{\bf x}_1 d^3{\bf p}_1 d^3{\bf x}_2 d^3{\bf p}_2 d^3{\bf x}_3 d^3{\bf p}_3 f_{q_1}({\bf x}_1, {\bf p}_1)\nonumber\\ 
&&\times f_{q_2}({\bf x}_2, {\bf p}_2)f_{q_3}({\bf x}_3, {\bf p}_3)W_{B}({\bf y}_{1},{\bf k}_{1};{\bf y}_{2},{\bf
k}_{2})\nonumber\\
&&\times \delta^{(3)}({\bf P}_{B}-{\bf p}_1 -{\bf p}_2 -{\bf p}_3) \, ,
\end{eqnarray}
respectively. In the above, $f_{q}({\bf x}_1,{\bf p}_1)$ and $f_{\bar{q}}({\bf x}_2, {\bf p}_2)$ are the phase-space distribution functions of quarks and antiquarks, and they are normalized as $\int d^3{\bf x}d^3{\bf p}f_{q, \bar q}({\bf x},{\bf  p})=N_{q,\bar q}$, where $N_{q,\bar q}$ is the quark or antiquary number. The Wigner functions of the meson and baryon (or antibaryon) are denoted by $W_{M}({\bf y},{\bf k})$ and $W_{B}({\bf y}_{1},{\bf k}_{1};{\bf y}_{2},{\bf k}_{2})$, expressed in terms of the relative coordinates and relative momenta of their valence quarks. For mesons, they are defined as
\begin{equation} \label{eq:rel}
{\bf y}={\bf x}_{1}-{\bf x}_{2},\quad {\bf k}=\frac{1}{m_1+m_2}(m_2{\bf p}_{1}-m_1{\bf p}_{2}),
\end{equation}
where $m_1$ and $m_2$ are the masses of the quark and antiquark, respectively. For baryons (or antibaryons), while ${\bf y}_1$ and ${\bf k}_1$ are similarly defined as in Eq.(\ref{eq:rel}), the second relative coordinate ${\bf y}_2$ and relative momentum ${\bf k}_2$
are given by
\begin{eqnarray}
\label{distB}
{\bf y}_2&=&\frac{m_1{\bf x}_1+m_2{\bf x}_2}{m_1+m_2}-{\bf x}_3,\nonumber\\
{\bf k}_2&=&\frac{m_3({\bf p}_1+{\bf p}_2)-(m_1+m_2){\bf p}_3}{m_1+m_2+m_3},
\end{eqnarray}
with $m_3$ being the mass of the third quark (or antiquark). The meson and baryon Wigner functions are normalized as 
$(2\pi)^{-3}\int d^3{\bf y}d^3{\bf k}f_M({\bf y},{\bf k})=1$ and
$(2\pi)^{-6}\int d^3{\bf y}_1d^3{\bf k}_1d^3{\bf y}_2d^3{\bf k}_2f_B({\bf y}_1,{\bf k}_1;{\bf y}_2,{\bf k}_2)=1$. 
The factor $g_{M}$ in Eq.(\ref{eq:rel}) accounts for the
probability for the color triplet, spin-$1/2$ quark and antiquark to form a
given color singlet meson, while $g_B$ is the corresponding factor for three
quarks (antiquarks) to form a given color singlet baryon (antibaryon).  
In the present study, the phase space functions of quarks and antiquarks will 
be replaced by the Wigner functions of individual quarks and antiquarks from the
Monte Carlo jet shower generator, and we are going to use harmonic oscillator wave functions for hadrons to evaluate the momentum and space-time overlap integrals. 

In Ref.~\cite{Greco:2003xt} for recombination of thermal partons among
themselves and with jet partons, both the color and spin quantum numbers are
treated on a purely statistical basis.  The color flow in the parton shower 
is in principle tractable, although not yet implemented here for simplicity.
Since the number of shower partons in a jet is very small, strong color 
correlations exist and the probability for colored shower partons to form 
color singlet hadrons is thus much larger than given by a statistical factor 
for colored thermal partons. For the present study we will neglect the 
statistical factors due to the color degrees of freedom and only include those due to the spin degrees of freedom. However, we prohibit the quark-antiquark pair from a forced gluon decay to recombine into a color-singlet meson.
This approximation can be solidified by either invoking local color neutrality
arguments~\cite{Amati:1978by,Corcella:2002jc}, as also used for cluster hadronization in HERWIG, or
a color octet approach, similar to the one used in heavy quarkonium production 
in nuclear reactions~\cite{Wong:1999ur} where color octet clusters are allowed
to exchange soft gluons to convert into color singlets. Of course this could
be improved in the future by following color flow in the parton shower simulation.

Because of their large relative momenta, shower partons are quite likely to
recombine into excited hadron states. Wigner functions can have negative
values, which makes them unsuitable for direct Monte Carlo
evaluation. Instead we have to sample the quantum mechanical overlap integrals 
of the hadron Wigner functions with the Wigner functions representing
the wave packets of shower partons, which we take to be Gaussians
here.  The resulting quantum mechanical overlap integral, which is guaranteed 
to provide a positive definite probability density that can be sampled, is equivalent to 
a Gaussian smearing of the Wigner functions in Eqs.(\ref{Eq1}) and (\ref{Eq2}), i.e., replacing
$W_M$ by 
\begin{eqnarray}
\label{Eq1a}
\overline W_{M}({\bf y},{\bf k})&=&\int d^3{\bf x}_1^\prime d^3{\bf k}_1^\prime
d^3{\bf x}_2^\prime d^3{\bf k}_2^\prime\nonumber\\  
&\times& W_q({\bf x}_1^\prime,{\bf k}_1^\prime)W_{\bar q}({\bf x}_2^\prime,{\bf k}_2^\prime)W_M({\bf y}^\prime,{\bf k}^\prime).
\end{eqnarray}
In the above, $W_q({\bf x}_1^\prime,{\bf k}_1^\prime)$ and $W_{\bar q}({\bf x}_2^\prime,{\bf k}_2^\prime)$ are, respectively, the Wigner  functions of the quark and antiquark with their centroids at $({\bf x}_1,{\bf k}_1)$ and $({\bf x}_2,{\bf k}_2)$, respectively.
The formula for baryons is analogous.
 
We can evaluate Eq.(\ref{Eq1a}) with the help of some mathematics worked out
in Appendix~\ref{wwwM}. The result for a meson in the $n$-th excited state in the center of mass frame of the 
quark-antiquark pair is 
\begin{eqnarray}\label{WigM3}
\overline W_{M,n} (\mathbf{y} ,\mathbf{k})=\frac{v^{n}}{n!}e^{-v}.
\end{eqnarray}
with
\begin{eqnarray}
\label{WigM}
v=\frac{1}{2}\left(\frac{{\bf y}^2}{\sigma_M^2}+{\bf k}^2 \sigma_M^2 \right).
\end{eqnarray} 
where $\sigma_M$ is the width of the harmonic oscillator wave function for the
relative motion of quark antiquark pair.

Similarly, the Gaussian smeared Wigner function for a baryon, with a wave function in the $n_1$-th excited state in one relative coordinate and in the $n_2$-th excited state in the other relative coordinate, is given by 
\begin{eqnarray}
\label{Wig3B}
\overline W_{B,n_1,n_2} (\mathbf{y}_1 , \mathbf{k}_1 ; \mathbf{y}_2, \mathbf{k}_2)=\frac{v_1^{n_1}}{n_1!}e^{-v_1}\cdot\frac{v_2^{n_2}}{n_2!}e^{-v_2},
\end{eqnarray}
with
\begin{eqnarray}
\label{WigM}
v_i=\frac{1}{2}\left(\frac{{\bf y}_{i} ^2}{\sigma_{Bi}^2}+{\bf k}_{i} ^2 \sigma_{B_i^2}\right),\quad i=1,2.
\end{eqnarray}

Since the wave functions of quarks and/or antiquarks in a hadron are always given in the rest frame of the hadron, we evaluate the relative coordinates and momenta in Eqs.(\ref{eq:rel}) and (\ref{distB}) using the parton coordinates and momenta given at \emph{constant rest
frame time}~\cite{Chen:2003ava,Chen:2006vc} in terms of their equal-time coordinates in the hadron rest frame. To this end, for each candidate partons to be treated their phase-space coordinates have to be Lorentz transformed from the lab frame to their common
rest frame, and subsequently the partons produced earlier in the parton shower are propagated like free particles to the time at which the last candidate parton is produced and available for hadronization. We have
checked that an algorithm that rather takes the distance of closest approach for the candidate partons has not much influence on the results as the parton shower is rapidly expanding. 

The two width parameters $\sigma_{B1}$ and $\sigma_{B2}$ in the baryon Wigner function are related to each other by 
\begin{eqnarray}
\label{BaryonSigma}
\sigma_{B2}=\sigma_{B1}\left(\frac{\mu_1}{\mu_2}\right)^{1/2},
\end{eqnarray}
where the two reduced masses are defined as \cite{Oh:2009zj} 
\begin{eqnarray}
\mu_1 =\frac{m_1 m_2}{m_1 +m_2},~~\mu_2 =\frac{(m_1 +m_2 )m_3}{m_1 +m_2 +m_3}.
\end{eqnarray}

The width parameters of the harmonic oscillator wave function can be related
to the measured size of the formed hadron.
More precisely, for a meson consisting of quark and antiquark of masses $m_1$ and $m_2$ and quark charges $Q_1$ and $Q_2$, its mean-square charge radius is related to $\sigma_M$ by~\cite{Oh:2009zj}
\begin{eqnarray}
\langle r^2 \rangle_M&=&|\langle Q_1({\bf x}_1-{\bf X})^2+Q_2({\bf x}_2-{\bf X})^2\rangle|\nonumber\\
&=&\frac{3}{2}\frac{|Q_1m_2^2+Q_2m_1^2|}{(m_1+m_2)^2}\sigma_M^2,
\end{eqnarray}
where ${\bf X}=(m_1{\bf x}_1+m_2{\bf x}_2)/(m_1+m_2)$ is the center of mass coordinate.

Similarly, the width parameter $\sigma_{B1}$ in the Wigner function of a baryon consisting of three quarks of masses $m_1$, $m_2$, and $m_3$, and charges $Q_1$, $Q_2$, and $Q_3$ are related to its mean-square charge radius is by~\cite{Song:2012cd}
\begin{widetext}
\begin{eqnarray}\label{radius}
\langle r^2 \rangle_B&=&|\langle Q_1({\bf x}_1-{\bf X})^2+Q_2({\bf x}_2-{\bf X})^2+Q_3({\bf x}_3-{\bf X})^2\rangle|\nonumber\\
&=&\frac{3\sigma_{B1} ^2}{2(m_1 +m_2 +m_3 )}\left[\frac{m_2 (m_2 +m_3 )}{m_1 +m_2}Q_1+\frac{m_1 (m_3 +m_1 )}{m_1 +m_2}Q_2 +\frac{m_1 +m_2}{m_3}Q_3 \right],
\end{eqnarray}
\end{widetext}
where ${\bf X}=(m_1{\bf x}_1+m_2{\bf x}_2+m_3{\bf x}_3)/(m_1+m_2+m_3)$ denotes the center of mass of the three quarks.

\begin{table}[t]\label{tab:1}
\caption{Empirical charge radii $R_c$ (from Ref.~\cite{Beringer:1900zz}), width parameters $\sigma_M$ or $\sigma_{B1}$, and spin statistical factors $g$ for hadrons used in the calculation.}
\vspace{0.3cm}
\begin{tabular}{c|ccc}
\hline\hline
Hadron & $R_c$ [fm] ~~& $\sigma_M$ or $\sigma_{B1}$ [fm] ~~& $g$ \\ \hline
$\pi$ & 0.67 & 1.09 & 1/4 \\ 
$\rho$ & -- & 1.09  & 3/4 \\ 
$K$ & 0.56 & 0.84  & 1/4 \\ 
$K^*$ & -- & 0.84  & 3/4 \\
$N$ & 0.88 & 1.24  &1/4 \\ 
$N^*$ & -- & 1.24 & 1/4 \\ 
$\Delta$ & -- & 1.24  & 1/2 \\ 
$\Lambda$ & -- & 1.21 & 1/4 \\ \hline\hline
\end{tabular}
\end{table}

We use measured charge radii for charged pions, protons and charged kaons to
determine the width parameters $\sigma_\pi$, $\sigma_K$ and $\sigma_N$ in the
pion, kaon and nucleon Wigner functions. The same width parameters are used
for their isospin partners and their antiparticles as well as their spin resonances
$\rho$, $K^*$, $N^*$, and $\Delta$.  Since $\Lambda$ and $\bar\Lambda$ have no
charge, their width parameters are determined instead from the matter radius,
which is given by an equation similar to Eq.\ (\ref{radius}) after setting
$Q_1=Q_2=Q_3=1/3$, and assuming that their size is the same as that of a
proton. Excited states of these hadrons are then accounted for by the excited
states of the harmonic oscillator wave functions using the same width
parameters.
In Table I we summarize the charge radii, width parameters and spin statistical factors for all stable hadrons and resonances included here.

Eqs.\ (\ref{WigM3}) and (\ref{Wig3B}) can now be used to determine the
recombination probability for a given quark-antiquark pair or a triplet of
quarks or antiquarks. For a given shower, the relative coordinates in the 
common rest frame are evaluated for all possible hadron candidates which are
subsequently accepted for recombination, or rejected, by Monte Carlo methods.

Some quarks might have quite small probabilities for recombination with any
other parton in the same shower.
In that case, there is a large probability that the Monte Carlo algorithm will
not find a recombination partner. Such quarks are typically far removed from
others in phase space, making all Wigner function overlap integrals small. The
reason for this to occur is the lack of confinement in what is essentially a
perturbative shower evolution. Of course isolated partons have to be connected
by strings to another color charge and Lund string fragmentation can take care 
of their hadronization. We deal with such partons far removed in phase space
by reconnecting them to other partons by QCD strings. This also includes undoing
the non-perturbative gluon splitting introduced earlier, if none of the
daughter quarks has found a recombination partner.
We thus form short strings of the types $(q,\bar{q})$ and 
$(q,g,g,\ldots,\bar{q})$ and then hand them over to PYTHIA 6.3 for 
hadronization.

\begin{figure}[tb]
\begin{center}
\includegraphics[width=\columnwidth]{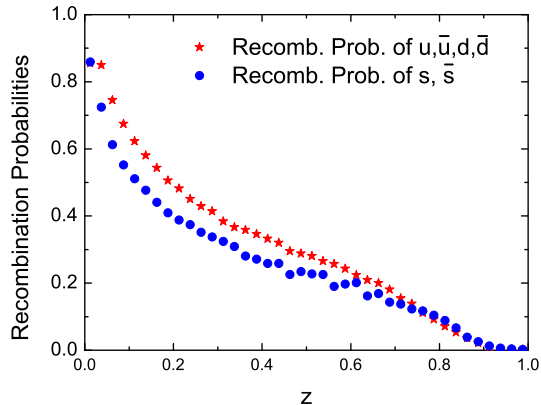}
\caption{(Color online) Probabilities of light and strange quarks to recombine
  into hadrons as functions of their momentum fraction $z$ for100 GeV jets.}
\label{probremn}
\end{center}
\end{figure}

We end up with the following picture: Final hadron spectra are a mixture of
hadrons from recombination (from quarks close in phase space to other quarks)
and from string fragmentation (for quarks isolated in phase space or otherwise
leftover). Typically, high-$z$ partons are both rare and far removed in phase
space. They are unlikely to recombine with other partons in the shower (or
partons from a surrounding medium if one would consider such). This can be
seen in Fig.\ \ref{probremn} where the probability of quarks to find a
recombination partner is plotted as a function of parton momentum fraction $z$
for 100 GeV jets. Thus in our model moderate to high-$z$ partons still
preferentially hadronize by string breaking. On the other hand, we indeed find
the existence of a bulk of jet partons at lower $z$ in which quarks are 
close enough in phase space so that they prefer recombination.
Recall that our main motivation is to establish a hadronization model which
naturally generalizes to jets in a medium. It is now straightforward to see how
our formalism can be applied to that more general case~\cite{Han:new}.

Excited states will be important channels for recombination. Excited mesons 
and baryons up to $n=5$  are known experimentally~\cite{Beringer:1900zz}. 
However, here we include the contributions from excited meson states up to $n=8$ 
and excited baryon states up to $n_1+n_2=8$, which can be easily
done with harmonic oscillator wave functions. We allow
excited states to decay to multiple pions in the case of light quark mesons,
to kaon and pion in the case of light and strange mesons, to (anti)nucleon and
pion in the case of light flavor (anti)baryons, and to $\Lambda$ and pion in
the case of strangeness $\pm 1$ baryons.  For decays into multiple pions, we
determine their relative probabilities through the available phase space according to~\cite{Milburn:1955zz}
\begin{eqnarray}
\label{prob}
P_l(M)\sim\left[\frac{1}{6\pi^2}\left(\frac{M}{m_\pi}\right)^3\right]^l\frac{(4l-4)!(2l-1)}{(2l-1)!^2(3l-4)!}
\, .
\end{eqnarray}
Here $l$ is the number of pions, $M$ is the mass of the excited state, or the invariant mass of the light quark-antiquark pair.
The pion mass $m_\pi$ in the above equation comes from taking the radius of
the emitting source to be that of the inverse of the pion
mass~\cite{Milburn:1955zz}.  In the present study, we replace $1/m_\pi$ by the
distance between the recombined quark and antiquark, and consider its decay to
at most four pions. The momentum distribution of these pions is then
determined from phase space considerations. An excited nucleon $N^*$ or
$\Delta$ decays to a nucleon and $l$ pions if its invariant mass is between
$m_N+lm_\pi$ and $m_N+(l+1)m_\pi$ with $m_N$ being the nucleon masses. Again,
we include at most four pions in the decay and use phase space considerations
to determine their momenta. An excited kaon or $\Lambda$ is assumed to decay
to a kaon or $\Lambda$ and multi-pions in a similar way.  

\section{Results}

In the following, we compare results from our hadronization model applied
to parton showers from PYTHIA 6.3 to calculations of PYTHIA with string
fragmentation applied to the same parton showers.

\begin{figure}[tb]
\begin{center}
\includegraphics[width=\columnwidth]{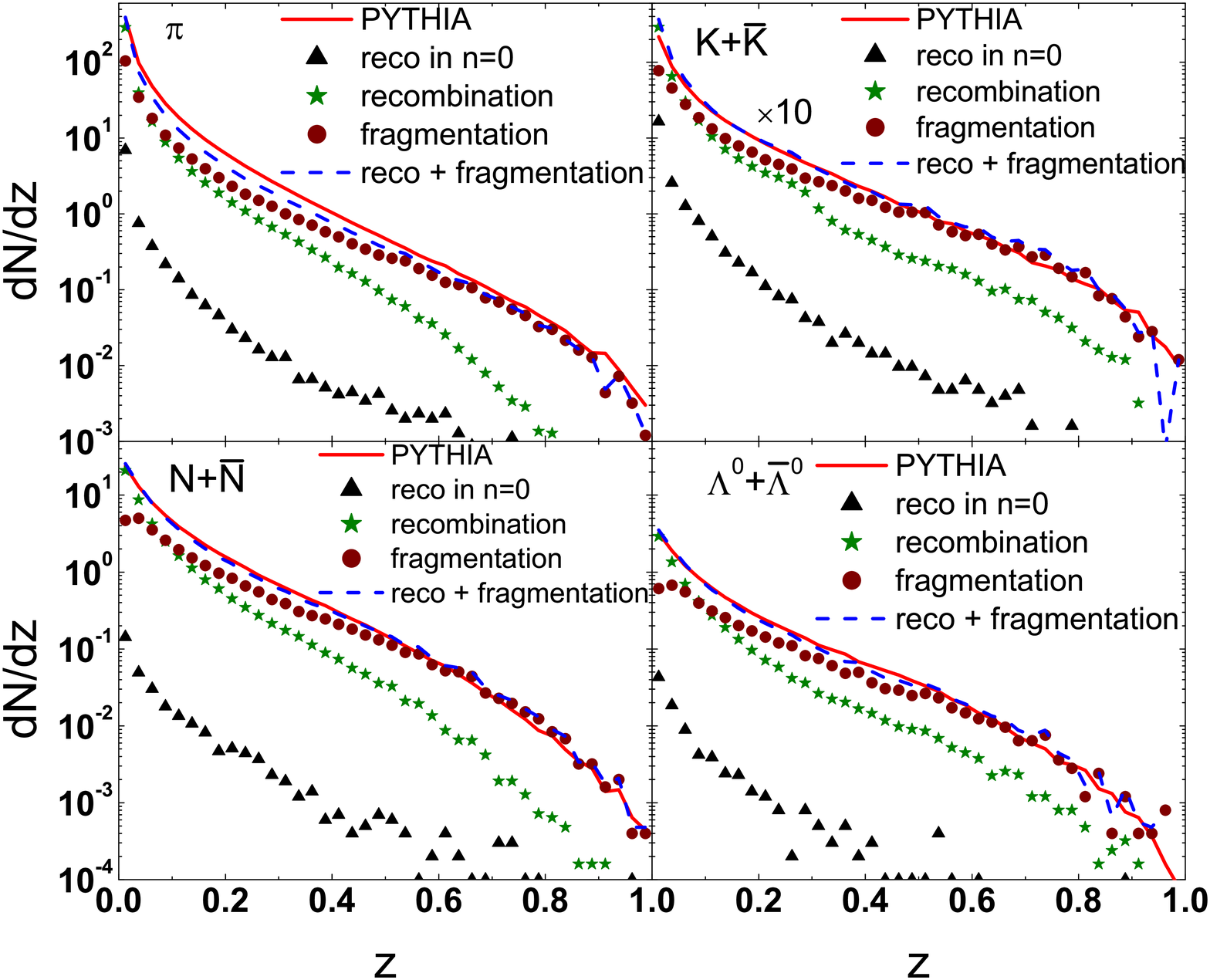}
\caption{(Color online) Spectrum $dN/dz$ of pions (upper left panel), kaons (upper right panel), nucleons and antinucleons (lower left panel), and $\Lambda$ and $\bar\Lambda$ (lower right panel) from our calculation. Shown separately are contributions from the recombination of shower partons (stars) and fragmentation of remnant partons (circles).  Also shown are the total contribution (dashed lines) and the results from PYTHIA string fragmentation (solid lines).}
\label{recomb}
\end{center}
\end{figure}

First, we test the longitudinal structure of jets by comparing the spectra
$dN/dz$ as functions of the momentum fraction $z$ longitudinal to the jet
axis for our sample of 100 GeV jets. In Fig.~\ref{recomb}, we show the spectra 
of pions (upper left panel),
kaons (upper right panel), nucleons and antinucleons (lower left panel), and
$\Lambda$ and $\bar\Lambda$ (lower right panel) from 100 GeV quark jets. 
We show separately hadrons from recombined shower partons (stars), from the 
fragmentation of remnant hadrons (circles) and their sum (dashed line).
The solid line indicates the result from PYTHIA 6.3 string fragmentation 
applied to the same sample of jet parton showers. We also show the recombination
only through the ground state of the harmonic oscillator wave functions 
($n=0$). As expected, we see that recombination spectra fall off faster with
$z$ than the string fragmentation contribution. String fragmentation dominates
at intermediate and high $z$ while recombination becomes the leading channel
below $z\sim 0.1$, where the bulk of the hadron production resides.

\begin{figure}[tb]
\begin{center}
\includegraphics[width=\columnwidth]{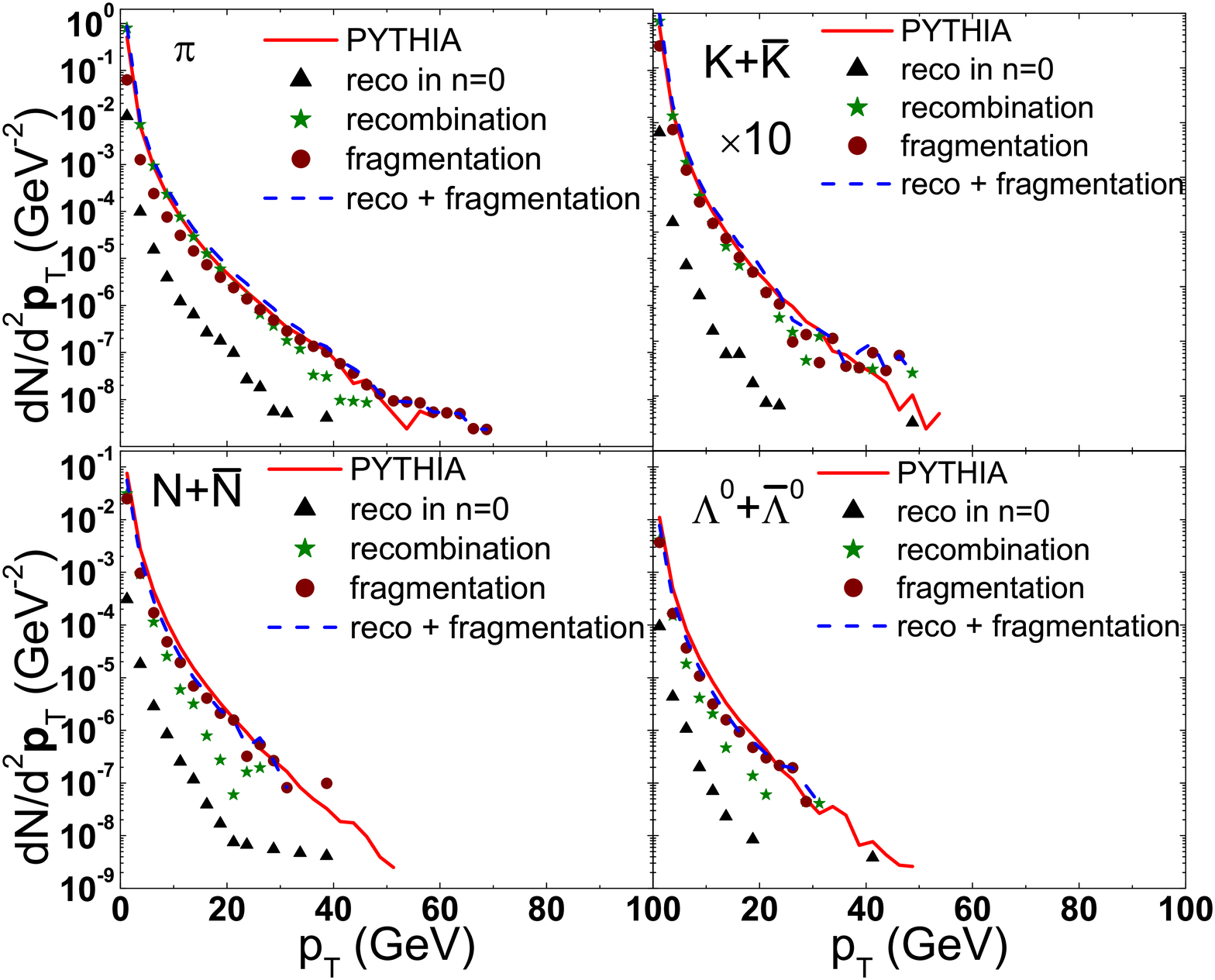}
\caption{(Color online) Same as Fig.~\ref{recomb} for the transverse momentum spectra.}
\label{recombpT}
\end{center}
\end{figure}

We also note that recombination proceeds mainly through excited hadron states
and not directly into $n=0$ ground state hadrons. The $n=0$ channel includes 
direct production of pion, kaon, nucleon, and $\Lambda$ as well as production 
from the the decay of $n=0$ spin-excited states $\rho$, $K^*$,
$N^*$ and $\Delta$. The inclusion of excited states $n>0$ makes the
recombination spectra considerably harder.
Overall we find that the results from our model are consistent with spectra
created by PYTHIA from pure string fragmentation.
Note that the comparison to string fragmentation --- another model --- only makes 
sense on a qualitative level. Precision tuning of our model would have to
involve fits to data which is outside the scope of this work.
We compare the transverse momentum spectra of jets in 
Fig.~\ref{recombpT}. Again, results obtained from our hybrid recombination
and fragmentation model compare well to pure string fragmentation.

\begin{figure}[tb]
\begin{center}
\includegraphics[width=\columnwidth]{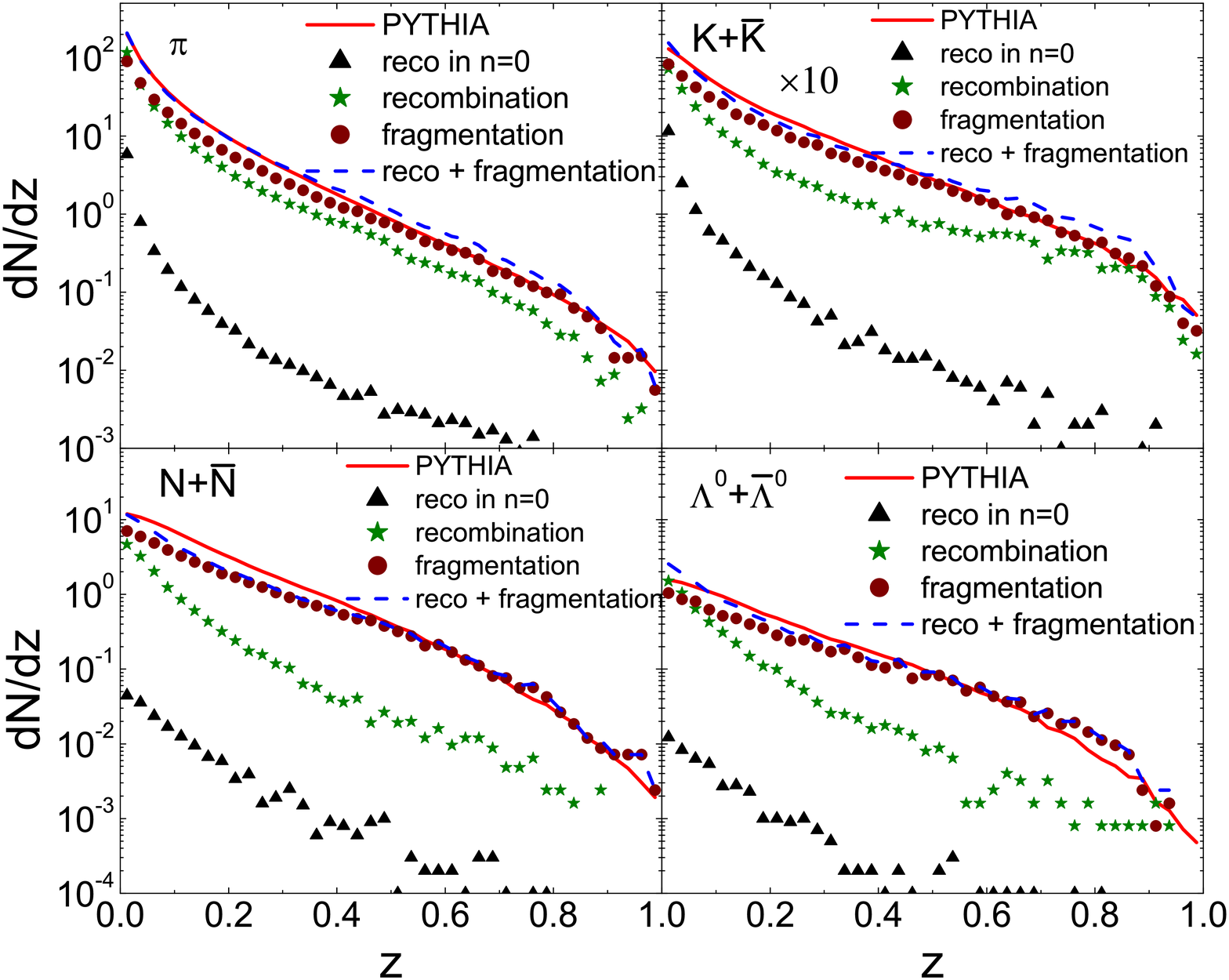}
\caption{(Color online) Same as Fig.~\ref{recomb} for jets of energy $E_\mathrm{jet}=25$ GeV.}
\label{recomb_50}
\end{center}
\end{figure}

\begin{figure}[tb]
\begin{center}
\includegraphics[width=\columnwidth]{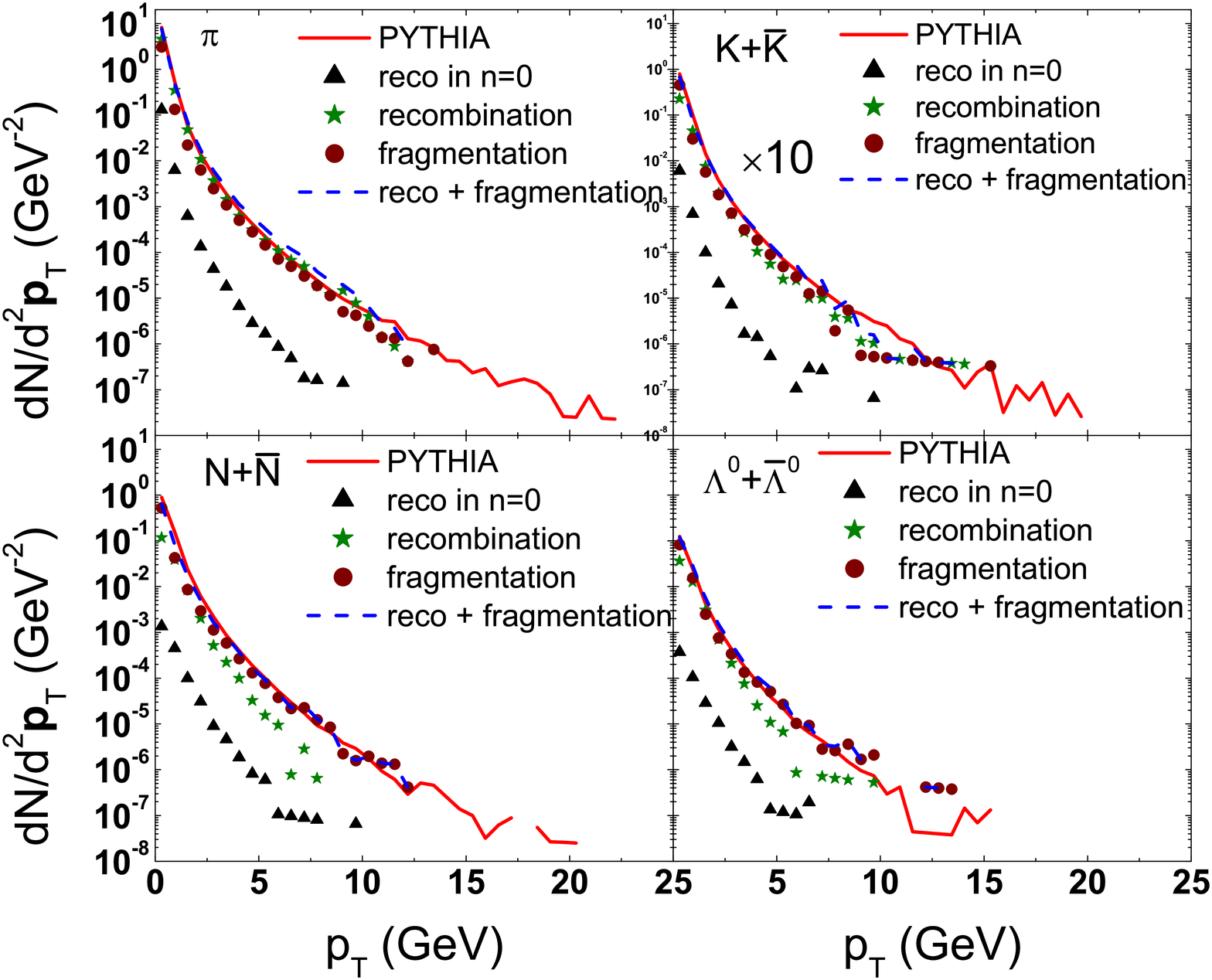}
\caption{(Color online) Same as Fig.~\ref{recombpT} for jets of energy $E_\mathrm{jet}=25$ GeV.}
\label{recomb_50pT}
\end{center}
\end{figure}

Finally we check our approach to hadronization with jets of a smaller jet
energy and find again that our results reproduce pure string fragmentation reasonably
well. The spectra for $E_\mathrm{jet}=25$ GeV jets are shown in 
Figs.~\ref{recomb_50} and \ref{recomb_50pT} for the longitudinal and
transverse momentum spectra, respectively.
The recombination probability depends on the absolute distance of partons
in phase space. Hence we expect the range in $z$ in which recombination
competes with remnant string fragmentation to decrease with rising $E_\mathrm{jet}$.
On the other hand, at smaller jet energies recombination stays more
competitive out to larger $z$ at least for mesons, while for baryons the
reduced number of partons in lower energy jets can lead to the opposite effect.

\section{Summary and Discussions}

We have devised a model to hadronize perturbative parton showers in jets based
on a hybrid of quark recombination and string fragmentation. Our algorithm
reproduces results from pure string fragmentation and can be easily generalized to include partons from an ambient medium.

We turn perturbative parton showers into showers of constituent quarks and antiquarks by gluon decay. We then apply Monte Carlo methods to recombine quarks and antiquarks using probabilities given by their overlap integrals with respect to meson and baryon Wigner functions. The width parameters in these Wigner functions are fixed by hadron charge radii. Remnant quarks and antiquarks, which are not used for recombination, are connected by strings and subjected to the usual string fragmentation procedure in PYTHIA. We find that decays of excited states from recombination make the most important contributions to spectra of pions, kaons and nucleons.

We have checked that both the longitudinal and transverse momentum structures of hadron showers reproduce the results from PYTHIA string fragmentation. The only adjustable parameter that we have kept is the mass cutoff for gluon decays into quark-antiquark pair which can be set by the strange to non-strange hadron ratio. However, other quantities which are not very well known, like the width parameters in the Wigner functions for excited states of hadrons, can in principle be used as parameters for further fine tuning of results.

Our hybrid approach essentially keeps string dynamics intact for the high-$z$
tail of the jet and replaces string dynamics with recombination for the bulk
of the jet where $\mathcal{O}(10)$ quarks with a few GeV/$c$ momentum can be found close enough together in phase space to recombine.

In the presence of a quark-gluon plasma produced in relativistic heavy ion collisions, we suggest that our approach  can be generalized by sampling the ambient medium (e.g.\ provided by a fluid dynamic simulation into which the jet is embedded) for thermal partons. Recombination would be delayed if the ambient temperature is above the critical temperature $T_c$.  At $T_c$ jet partons would be allowed to recombine with thermal partons, and remnant jet partons could also be allowed to connect to thermal partons by strings. This process, like other jet-medium interactions, would allow the exchange of energy and momentum. Details of an in-medium algorithm will be provided in a forthcoming manuscript \cite{Han:new}.

\section*{Acknowledgments}

We thank the members of the Recombination Working Group of the JET
collaboration for helpful discussions. This work was supported by the
U.S. National Science Foundation under Grants PHY-0847538, PHY-1516590 and
PHY-1068572, by the US Department of Energy under Contract
No. DE-FG02-10ER41682 within the framework of the JET Collaboration, and by the Welch Foundation under Grant No. A-1358.

\appendix*
\section{The Overlap of Gaussian Wave Packets with a Harmonic Oscillator}
\label{wwwM}

In this Appendix, we discuss details of the calculation of the overlap of
harmonic oscillator Wigner functions and Gaussian wave packets.
Let us start by noting that both Gaussian wave packets and the
harmonic oscillator problem factorize into the three spatial directions. We
can thus solve the corresponding one dimensional problem and then readily 
find the solution in three dimensions.

We start with the well known harmonic oscillator basis\
in one dimension~\cite{MerzQM},  
\begin{eqnarray}
\phi_{n}(x)=\left(\frac{m\omega}{\pi\hbar}\right)^{1/4}\frac{1}{\sqrt{2^n n!}}H_{n}(\xi)e^{-\xi^2 /2},
\end{eqnarray}
where $\xi=\sqrt{\frac{m\omega}{\hbar}}x$, $H_{n}(\xi)$ are Hermite
polynomials and $\omega$ is the oscillator frequency.
The Wigner transformation of the harmonic oscillator wave functions, defined by
\begin{eqnarray}
W_{n}(x,k )=\int^{\infty} _{-\infty} d\eta~ e^{i k \eta}\phi_n \left(x+\frac{\eta}{2}\right)\phi_n \left(x+\frac{\eta}{2}\right),
\end{eqnarray}
leads to~\cite{Groenewold:1946kp} 
\begin{eqnarray}\label{lag}
W_n (u)= 2(-1)^n L_{n} (u)e^{-u/2},
\end{eqnarray}
where $u=2\left(\frac{x^2}{\sigma^2}+\sigma^2 k^2\right)$ with the width
$\sigma=\left(\frac{\hbar}{m\omega}\right)^{1/2}$, and the $L_n$ are Laguerre polynomials.

We would like to calculate the overlap integral
\begin{eqnarray}\label{recopack}
\overline W_n (x, k)&=&\int dx_1^\prime dk_1^\prime dx_2^\prime dk_2^\prime\nonumber\\
&\times& W(x_1^\prime,k_1^\prime)W(x_2^\prime,k_2^\prime)W_n (x,k),
\end{eqnarray}
of the Wigner function with Gaussian wave packets
\begin{eqnarray}
\label{wavepacket}
W(x_i^\prime ,k_i^\prime)=\frac{1}{\pi}e^{-(x_i^\prime -x_{i} )^2 /\delta^2} e^{-\delta^2(k_i^\prime-k_{i} )^2},~i=1,2,\nonumber\\
\end{eqnarray}
of width $\delta$ around centroids $x_{i}$ and $k_{i}$ in space and momentum. Here $x=x_1-x_2$ and $k = (k_{1}-k_{2})/{2}$, and the result will only depend on the relative position of centroids $r=x_{1} -x_{2}$ and 
$p =(k_{1}-k_{2})/{2}$.

Using the generating function for Laguerre polynomials~\cite{MathMPhy}, 
\begin{eqnarray}
\label{genf}
\frac{1}{1-t}e^{-\frac{tx}{1-t}}=\sum^{\infty} _{n=0} t^n L_n (x),
\end{eqnarray}
it is straightforward to see that Eq.\ (\ref{lag}) leads to the generating function
for the oscillator Wigner functions
\begin{eqnarray}
\frac{2}{1+t}\exp\left(-\frac{1-t}{2(1+t)}u\right)=\sum^{\infty} _{n=0} t^n W_n (u).
\end{eqnarray}
Carrying out the integrals from Eq.\ (\ref{recopack}) on both sides of above
equation, we obtain the following generating function for the Gaussian smeared
Wigner function $\overline W_n$ 
\begin{eqnarray}
&&\frac{2}{(1+t)(1+a\alpha)^{1/2}(1+a\alpha^{-1})^{1/2}}\nonumber\\
&&\times\exp\left(-\frac{ax^2}{(1+a\alpha)\sigma^2}-\frac{ak^2 \sigma^2}{1+a\alpha^{-1}}\right)=\sum^{\infty} _{n=0}t^n \overline W_n (x,k),
\nonumber\\
\end{eqnarray}
where $a=\frac{1-t}{1+t}$ and $\alpha=2\delta^2/\sigma^2$.  By Taylor
expanding the left hand side in $t$ and comparing coefficients of the same powers in $t$ on both sides, we obtain the following recurrence relation for the $\overline W_n$:
\begin{eqnarray}
\label{recurM}
\overline W_{n+5}&=&-\frac{1}{\Lambda_5}(\Lambda_4\overline W_{n+4}+\Lambda_3\overline W_{n+3}+\Lambda_2\overline W_{n+2} \nonumber\\
&&+\Lambda_1\overline W_{n+1}+\Lambda_0\overline W_n),
\end{eqnarray}
where $\Lambda_{i}~(i=0,1,\ldots,5)$ are given by
\begin{eqnarray}
\Lambda_0 &=&-[(1+\alpha)^2 +n](1-\alpha)^2,\nonumber\\
\Lambda_1 &=&[\alpha(1-\alpha)+2(x/\sigma)^2 +2\alpha^2 (k\sigma)^2+n+1](1-\alpha)^2 x \nonumber\\
\Lambda_2 &=&[(1-\alpha)(\alpha^2 +4\alpha+1)-2(x/\sigma)^2 (3\alpha +1)\nonumber \\
&&-2\alpha(k\sigma)^2(-\alpha^2 +3\alpha+2)\nonumber\\
&&-2(n+2)(1+\alpha)^2(1-\alpha)](1-\alpha),\nonumber\\
\Lambda_3 &=&[\alpha(1-\alpha)^2 +2(x/\sigma)^2 (3\alpha^2 -2\alpha-1)\nonumber \\
&&+2\alpha(k\sigma)^2 (\alpha^3 -3\alpha^2 +9\alpha-7)\nonumber \\
&&-2(n+3)(1+\alpha)^2 (1-\alpha)^2 ], \nonumber\\
\Lambda_4 &=&[2(x/\sigma)^2 +2\alpha^2 (k\sigma)^2 -(n+4)(1-\alpha)^2](1+\alpha)^2,\nonumber\\
\Lambda_5 &=&-(n+5)(1+\alpha)^2.
\end{eqnarray}

Taking $\alpha=2\delta^2 /\sigma^2 = 1$ for simplicity reduces Eq.(\ref{recurM}) to 
\begin{eqnarray}\label{rec}
\overline W_{n+1} =\frac{v}{n+1}\overline W_n,
\end{eqnarray}
with 
\begin{eqnarray}
\overline W_0 =\exp(-v) ~~{\rm and}~~ v=\frac{1}{2}\left(\frac{x^2}{\sigma^2}+k^2 \sigma^2\right)
\end{eqnarray}
or equivalently 
\begin{eqnarray}
\overline W_n = \frac{v^n}{n!}e^{-v}.
\end{eqnarray}
The Gaussian smeared $\overline W_n$ has the form of a Poisson distribution with the normalization
\begin{eqnarray}
\sum^{\infty} _{n=0} \frac{v^n}{n!}e^{-v}=1,
\end{eqnarray}
which is similar to that of a coherent state~\cite{sakurai}.

In three dimensions, the Gaussian smeared Wigner function is thus given by 
\begin{eqnarray}\label{WigM3}
&&\overline W_{n} (\mathbf{x} ,\mathbf{k})\nonumber\\
&&=\sum_{n_{x}+n_{y}+n_{z}=n}\overline W_{n_{x}} (x,k_{x} )\overline W_{n_{y}} (y,k_{y} )\overline W_{n_{z}} (z,k_{z} )\nonumber\\
&&=\sum_{n_{x}+n_{y}+n_{z}=n}\frac{v_x^{n_x}}{n_{x}!}e^{-v_x}\cdot\frac{v_y^{n_x}}{n_{y}!}e^{-v_y}\cdot\frac{v_z^{n_z}}{n_{z}!}e^{-v_z}\nonumber\\
&&=\frac{1}{n!}e^{-v}\sum_{n_{x}+n_{y}+n_{z}=n}\frac{n!}{n_x!n_y!n_z!}v_x^{n_x}v_y^{n_y}v_z^{n_z}\nonumber\\
&&=\frac{v^n}{n!}e^{-v},
\end{eqnarray}
with
\begin{eqnarray}
\label{WigM}
v=v_{x}+v_{y}+v_{z}=\frac{1}{2}\left(\frac{{\bf x} ^2}{\sigma^2}+{\bf k}^2 \sigma^2 \right).
\end{eqnarray}
The fourth equality in Eq.(\ref{WigM3}) follows from using the trinomial expansion formula 
\begin{eqnarray}
\sum_{n_{x}+n_{y}+n_{z}=n}&&\frac{n!}{n_x!n_y!n_z!}v_x^{n_x}v_y^{n_y}v_z^{n_z}\nonumber\\
&&=(v_x+v_y+v_z)^n=v^n.
\end{eqnarray}

\end{document}